\documentclass{svmult}

\usepackage{url}             
\usepackage{times}           
\usepackage{makeidx}         
\usepackage{graphicx}        
\usepackage{multicol}        
\usepackage[bottom]{footmisc}

\makeindex             

\begin{document}

\title*{Market Polarization in Presence of Individual Choice Volatility}
\author{Sitabhra Sinha\inst{1}\and
Srinivas Raghavendra\inst{2}}
\institute{The Institute of Mathematical Sciences, CIT Campus,
Taramani, Chennai 600113, India
\texttt{sitabhra@imsc.res.in}
\and Department of Economics, National University of Ireland, Galway, Ireland \texttt{s.raghav@nuigalway.ie}}
%
%
\maketitle

Financial markets are subject to long periods of polarized behavior,
such as bull-market or bear-market phases, in which the vast majority
of market participants seem to almost exclusively choose one action
(between buying or selling) over the other. From the point of view of
conventional economic theory, such events are thought to reflect the 
arrival of ``external news'' that justifies the observed behavior. 
However, empirical observations of the events leading up to such market 
phases, as well events occurring during the lifetime of such a phase, have 
often failed to find significant correlation between news from outside 
the market and the behavior of the agents comprising the market.
In this paper, we explore the alternative hypothesis that the occurrence
of such market polarizations are due to interactions amongst the agents
in the market, and not due to any influence external to it. In
particular, we present a model where the market (i.e., the aggregate behavior
of all the agents) is observed to become polarized even though
individual agents regularly change their actions (buy or sell) on
a time-scale much shorter than that of the market polarization phase.

\section{Introduction}
The past decade has seen an influx of ideas and techniques from physics
into economics and other social sciences, prompting some to dub this
new interdisciplinary venture as ``econophysics'' \cite{farmer05}. However,
it is not just physicists who have migrated to working on problems in such
non-traditional areas; social scientists have also started to use 
tools from, e.g., statistical mechanics, for understanding
various socioeconomic phenomena as the outcomes of interactions between
{\em agents}, which may represent individuals, firms or nations 
(see for example, Ref.~\cite{durlauf99}). 
The behavior of financial markets, in particular, has become a focus
of this kind of multidisciplinary research, partly because of the 
large amount of empirical data available for such systems. This makes
it possible to construct quantitatively predictive theories for such
systems, and their subsequent validation. 

Analysis of the empirical data from different financial markets has led to the
discovery of several {\em stylized facts}, i.e., features that are relatively
invariant with respect to the particular market under study. For example,
it seems to be the case that markets (regardless of their stage of development)
show much stronger fluctuations than
would be expected from a purely Gaussian process \cite{gopikrishnan98,sinha06a}.
Another phenomenon that has been widely reported in financial markets 
is the existence of {\em polarized} phases, 
when the majority of market participants
seem to opt exclusively to buy rather than sell (or vice versa) for prolonged
periods. Such bull-market (or bear-market) phases, when the market 
exhibits excess demand (or supply) relative to the market {\em equilibrium} 
state,
where the demand and supply are assumed to balance each other, are quite
common and may be of substantial duration. Such events are less spectacular
than episodes of speculative bubbles and crashes \cite{schiller00}, 
which occur over a 
relatively faster time-scale; however, their impact on the general economic
development of nations maybe quite significant, partly because of their
prolonged nature. Hence, it is important to
understand the reasons for occurrence of such market polarizations.

Conventional economic theory seeks to explain such events as reflections
of news external to the market. If it is indeed true that particular
episodes of market polarizations can only be understood as
responses to specific historical contingencies, then it should be
possible to identify the significant historical events that precipitated
each polarized phase. However, although {\em a posteriori} explanation
of any particular event is always possible, there does not seem to be 
any general explanation for such events in terms of extra-market variables,
especially one that can be used to predict future market phases.

In contrast to this preceding approach, one can view the market behavior
entirely as an emergent outcome of the interactions between the agents
comprising the market. While external factors may indeed influence
the actions of such agents, and hence the market, they are no longer
the main determinants of market dynamics, and it should be possible
to observe the various ``stylized facts'' even in the absence of
news from outside the market. In this explanatory framework, the
occurrence of market polarization can be understood in terms of time
evolution of the collective action of agents. It is important to
note here that the individual agents are assumed to exercise
their free will in choosing their particular course of action (i.e., whether
to buy or sell). 
However, in any real-life situation, an agent's action is also
determined by the information it has access to about the possible consequences
of the alternative choices available to it.
In a free market economy, devoid of any central coordinating authority, 
the personal information available to each agent may be different.
Thus the emergence of market behavior, which is a reflection of the collective 
action of agents, can be viewed as a
self-organized coordination phenomenon in a system of heterogeneous entities.

The simplest model of collective action is one where the action of
each agent is completely independent of the others; in other words, agents
choose from the available alternatives at random. In the case of
binary choice, where only two options are available to each agent, it is
easy to see that the emergence of collective action is equivalent to
a random walk on a one-dimensional line, with the number of steps equal to the
number of agents. Therefore, the result will be a Gaussian distribution,
with the most probable outcome being an equal number of agents choosing
each alternative. As a result, for most of the time the market will
be balanced, with neither excess demand nor supply. As already mentioned,
while this would indeed be expected in the idealised situation of
conventional economic theory, it is contrary to observations in
real life indicating strongly polarized collective behavior among agents
in a market. In these cases, a significant majority of agents choose one 
alternative over another, resulting in the market being either
in a buying or selling phase. Examples of such strong bimodal behavior has 
been also observed in contexts other than financial markets, e.g., in
the distribution of opening gross income for movies released in theaters
across the USA \cite{sinha04b}.

The polarization of collective action suggests that the agents do not
choose their course of action completely independently, but are
influenced by neighboring agents.
In addition, their personal information may change over time as a
result of the outcome of their previous choices, e.g., whether or not
their choice of action agreed with that of the majority
\footnote{This would be the case if, as in Keynes' ``beauty contest'' analogy
for the stock market, agents are more interested in foreseeing how
the general public will value certain investments in the immediate future,
rather than the long-term probable yields of these investments
based on their fundamental value \cite{keynes34}.}. This latter effect
is an example of global feedback process that we think is crucial
for the polarization of the collective action of agents, and hence, the market.

In this paper, we propose a model for the dynamics of market behavior
which takes into account these different effects in the decision process of
an agent choosing between two alternatives (e.g., buy or sell)
at any given time instant.
We observe a phase transition in the market behavior from an equilibrium 
state to a far-from-equilibrium state characterized by either excess demand
or excess supply under various conditions. However, most strikingly,
we observe that the transition to polarized market states occurs
when an agent learns to adjust its action according to
whether or not its previous choice accorded with that of the majority.
One of the striking consequences of this global feedback is that, although
individual agents continue to regularly switch between the alternatives
available to it, 
the duration of the polarized phase (during which the collective action
is dominated by one of the alternatives) can become extremely long.
The rest of the paper is organized as follows. In the next section, we 
give a detailed description of the model, followed in the subsequent
section by a summary of the results. We conclude with a discussion
of possible extensions of the model and implications of our results.
For further details please refer to Ref.~\cite{sinha04}.

\section{The Model}
\begin{figure}[tbp]
\centering
\includegraphics[height=6cm]{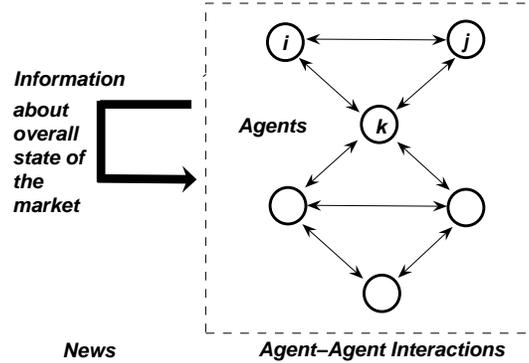}
\caption{An abstract model of a market. Each agent interacts (interactions
indicated by arrows) with a
subset of the other agents comprising the market, indicated by the boundary
formed from the broken lines. The combined action of all agents 
results in the overall
state of the market. The news of this state is 
available to all agents, although the information about the
individual actions of all agents 
may not be accessible to any one agent. 
}
\label{fig:1}       
\end{figure}
In this section we present a general model of collective action that
shows how polarization in the presence of individual choice volatility
can be achieved through adaptation and learning.
We assume that individual agents behave in a rational manner,
where rationality is identified with actions that would result in
market equilibrium in the absence of interaction between agents.
Therefore, for a large ensemble of such non-interacting agents we
will observe only small fluctuations about the equilibrium. Here we
explore how the situation alters when agents are allowed to 
interact with each other.
In our model, the market
behavior reflects the collective action of many interacting agents, 
each deciding to buy or sell based on limited information
available to it about the consequences of such action. 
An example of such limited information available
to an agent is news of the overall market sentiment as reflected
in market indices such as S~\&~P 500. A schematic diagram of the
various influences acting in the market is shown in Fig.~\ref{fig:1}.

Our model is defined as follows.
Consider a population of $N$ agents, whose actions are subject to
bounded rationality, i.e., they either buy or sell an
asset based on information about the action of their neighboring agents and
how successful their previous actions were. The fundamental value of the asset
is assumed to be unchanged throughout the period. 
In addition, the agents are assumed to have limited
resources, so that they cannot continue to buy or sell indefinitely. However,
instead of introducing explicit budget constraints \cite{iori02}, 
we have implemented
gradually diminishing returns for a decision that is taken repeatedly.
This is akin to the belief adaptation process in the Weisbuch-Stauffer
model of social percolation \cite{weisbuch03}, where making similar choices in
successive periods decreases the probability of making the same choice
in the subsequent period.

At any given time $t$, the state of an agent $i$ is fully described by two 
variables: its choice, $S^t_i$, and its belief about the outcome of 
the choice, $\theta^t_i$.
The choice can be either {\em buy}
($= +1$) or {\em sell} ($= -1$), while the belief can vary continuously 
over a range (initially, it is chosen from a uniform random distribution).
At each time step, every agent considers the average choice of its
neighbors at the previous instant, and if this exceeds its belief, then it
makes the same choice; otherwise, it makes the opposite choice.
Then, for the $i$-th agent, the choice dynamics is described by:
\begin{equation}
S_i^{t+1} = {\rm sign} ( \Sigma_{j \in {\cal N}} J_{ij} S_j^t - \theta_i^t),
\label{seqn}
\end{equation}
where $\cal N$ is the set of neighbors of agent $i$ ($i = 1, \ldots, N$),
and sign ($z$) = $+1$, if $z > 0$, and = $-1$, otherwise.
The degree of interaction among neighboring agents, $J_{ij}$, is assumed to 
be a constant ($= 1$) for simplicity and normalized by $z$ ($= |{\cal N}|$), 
the number of
neighbors. In a lattice, ${\cal N}$ is the set of spatial nearest neighbors and
$z$ is the coordination number,
while in the mean field approximation, ${\cal N}$
is the set of all other agents in the system
and $z = N - 1$.

The individual belief, $\theta$ evolves over time as:
\begin{equation}
\theta_i^{t+1} = \left\{ \begin{array}{ll}
\theta_i^t + \mu S_i^{t+1} + \lambda S_i^t, & {\rm if}~
S_i^{t} \neq {\rm sign}(M^t), \\
\theta_i^t + \mu S_i^{t+1}, & \mbox{otherwise,}
\end{array} \right.
\label{thetaeqn}
\end{equation}
where $M^t = (1/N) \Sigma_j S_j^t$ is the fractional excess demand and
describes the overall state of the market at any given time $t$. The
adaptation rate $\mu$ governs the time-scale of diminishing returns, over
which the agent switches from one choice to another in the absence of any
interactions between agents. The learning rate $\lambda$ controls the
process by which an agent's belief is modified when its action does not agree 
with that of the majority at the previous instant. As mentioned earlier,
the desirability of a particular
choice is assumed to be related to the fraction of the community choosing
it. Hence, at any given time, every agent is trying to coordinate its
choice with that of the majority. Note that, for $\mu = 0, \lambda = 0$,
the model reduces to the well-known zero-temperature, random field Ising model
(RFIM) of statistical physics.

We have also considered a 3-state model, where, in addition to $\pm 1$,
$S^t_i$ has a third state, $0$, which corresponds to the agent choosing
neither to buy nor sell. The corresponding choice dynamics, Eq.~(\ref{seqn}),
is suitably modified by introducing a threshold, with the choice variable
taking a finite value only if the magnitude of the difference between
the average choice of its neighbors and its belief exceeds this threshold. 
This is possibly a more realistic model of markets where an agent may choose
not to trade, rather than making a choice only between buying or selling.
However, as the results are qualitatively almost identical to the 2-state
model introduced before, in the following section we shall confine our
discussion to the latter model only.
\section{Results}
In this section, we report the main results of the 2-state model 
introduced in the preceding section. As the connection topology of
the contact network of agents is not known, we consider both the
case where the agents are connected to each other at random,
as well as, the case where agents are connected only to agents who
are located at spatially neighboring locations. Both situations are idealised,
and in reality is likely to be somewhere in between. However, it is
significant that in both of these very different situations we observe
market polarization phases which are of much longer duration compared
to the timescale at which the individual agents switch their choice
state ($S$).
\subsection{Random network of agents and the mean field model}
We choose the $z$ neighbors of an agent at
random from the $N-1$ other agents in the system. We also assume this
randomness to be ``annealed'', i.e., the next time the same agent
interacts with $z$ other agents, they are chosen at random anew.
Thus, by ignoring spatial correlations, a mean field approximation is
achieved.

For $z = N - 1$, i.e., when every agent has the information about the
entire system, it is easy to see that,
in the absence of learning ($\lambda = 0$),
the collective decision $M$
follows the evolution equation rule:
\begin{equation}
M^{t+1} = {\rm sign} [ (1-\mu) M^t - \mu \Sigma_{\tau = 1}^{t-1} M^{\tau}].
\end{equation}
For $0 < \mu < 1$, the system alternates between the states
$M=\pm 1$ (i.e., every agent is a buyer, or every agent is a seller)
with a period $\sim
4/\mu$. The residence time at any one state ($\sim 2/\mu$) increases with
decreasing $\mu$, and for $\mu = 0$, the system remains fixed at one of the
states corresponding to $M=\pm 1$, as expected from RFIM results.
At $\mu = 1$, the system remains in the market equilibrium state (i.e., $M = 0$).
Therefore, we see a transition from a bimodal distribution of the fractional
excess demand, $M$, with peaks at non-zero values, to an unimodal distribution
of $M$ centered about 0, at $\mu_c = 1$. When we introduce learning,
so that $\lambda > 0$, the agents try to coordinate with each other
and at the limit $\lambda \rightarrow \infty$ it is easy to see that
$S_i = {\rm sign}(M)$ for all $i$, so that all the agents make identical choice.
In the simulations, we note that the bimodal distribution is recovered
for $\mu =1$ when $\lambda \geq 1$.

For finite values of $z$, the population is no longer ``well-mixed''
and the mean-field approximation becomes less accurate the lower $z$ is.
For $z < < N$, the critical value of $\mu$ at which the transition
from a bimodal to a unimodal distribution occurs in the absence of
learning, $\mu_c < 1$. For example,
$\mu_c = 0$ for $z = 2$, while it is 3/4 for $z = 4$. As $z$ increases,
$\mu_c$ quickly converges to the mean-field value, $\mu_c = 1$.
On introducing learning
($\lambda > 0$)
for $\mu > \mu_c$, we again notice a transition to a state corresponding
to all agents being buyers (or all agents being sellers),
with more and more agents coordinating their choice.

\subsection{Agents on a spatial lattice}
To implement the model when the neighbors are spatially
related, we consider $d$-dimensional lattices ($d = 1, 2, 3$) and
study the dynamics numerically. We report results obtained in systems with
absorbing boundary conditions; using periodic boundary conditions leads to
minor changes
but the overall qualitative results remain the same.

\begin{figure}[tbp]
\centering
\includegraphics[height=12cm]{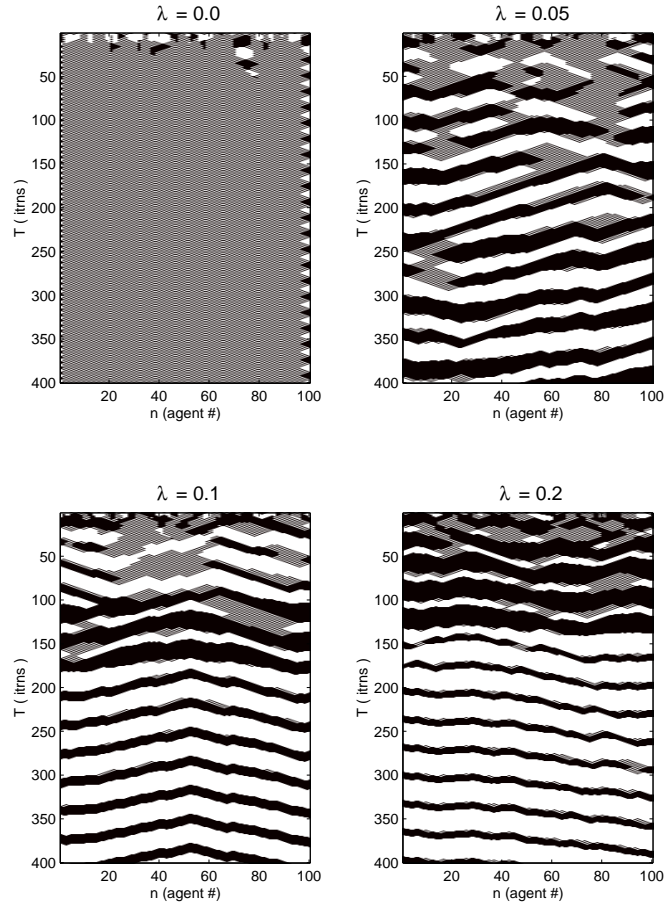}
\caption{The spatiotemporal evolution of choice ($S$) among $100$ agents,
arranged in a one-dimensional lattice, with the time-evolution upto $400$
iterations starting from a random configuration shown along the vertical
axis. The colors (white or black) represent the different choice states
(buy or sell) of individual agents.
The adaptation rate $\mu = 0.1$, and the learning rate $\lambda$ 
increases from 0~(top left) to 0.2~(bottom right). Note that, as
$\lambda$ increases, one of the two states becomes dominant with the majority 
of agents at any given time always belonging to this state,
although each agent regularly switches between the two states.
}
\label{fig:4}
\end{figure}

\begin{figure}[tbp]
\centering
\includegraphics[height=8cm]{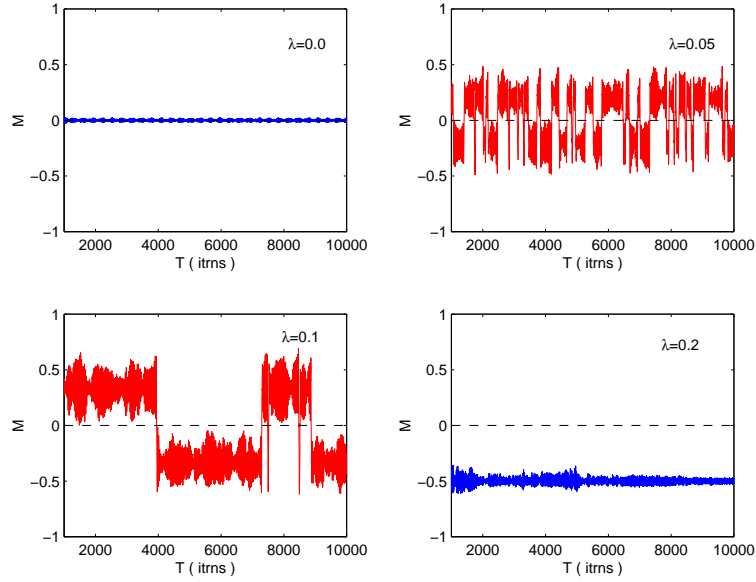}
\caption{Time series of the fractional excess demand $M$ in a
two-dimensional square lattice of $100 \times 100$ agents. The
adaptation rate $\mu = 0.1$, and the learning rate $\lambda$ is increased
from 0 to 0.2 to show the divergence of the residence time of the system
in polarized
configurations.
}
\label{fig:3}       
\end{figure}

\begin{figure}[tbp]
\centering
\includegraphics[height=6cm]{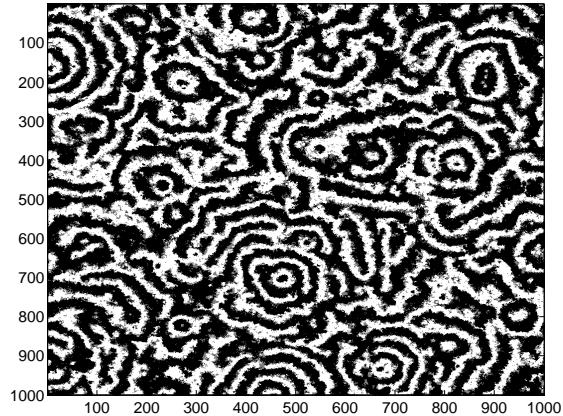}
\caption{The spatial pattern of choice ($S$) in a
two-dimensional square lattice of $100 \times 100$ agents after $2 \times 10^4$
iterations starting from a random configuration. The
adaptation rate $\mu = 0.1$, and the learning rate $\lambda$  of each agent
is randomly chosen from an uniform distribution between 0 and 0.1.
}
\label{fig:2}       
\end{figure}

In the absence of learning ($\lambda = 0$), starting from an initial
random distribution of choices and beliefs,
we observe only very small
clusters of similar choice behavior
and the fractional excess demand, $M$, fluctuates
around 0. In other words, at any given time an equal number
of agents (on average) make opposite choices so that the demand and
supply are balanced. In fact,
the most stable state under this condition is one where neighboring agents 
in the lattice make opposite choices.
This manifests itself as a checkerboard pattern in simulations carried
out in one- and two-dimensional square lattices (see e.g.,
Fig.~\ref{fig:4}, top left).
Introduction of
learning in the model ($\lambda > 0$) gives rise to significant clustering
among the choice of neighboring agents (Fig.~\ref{fig:4}), 
as well as, a large non-zero value for 
the fractional excess demand, $M$. 
We find that
the probability distribution of $M$ evolves from a single
peak at 0, to a bimodal distribution (having two peaks at finite values
of $M$, symmetrically located about 0) as $\lambda$ increases from 0
\cite{sinha05}.
The fractional excess demand
switches periodically from a positive value to a negative value
having an average residence time which increases sharply with
$\lambda$ and with $N$ (Fig. \ref{fig:3}).
For instance, when $\lambda$ is very high relative to $\mu$, we see
that $M$ gets locked into one of two states
(depending on the initial condition),
corresponding to the majority preferring either one or the other
choice. This is reminiscent of {\em lock-in} 
in certain economic systems subject
to positive feedback \cite{arthur89}. The special case
of $\mu = 0, \lambda > 0$ also results in a lock-in of the
fractional excess demand, with the time required to get to this state
increasing rapidly as $\lambda \rightarrow 0$.
For $\mu > \lambda > 0$, large clusters of agents with identical choice
are observed to form and dissipate throughout the lattice. After
sufficiently long times, we observe the emergence of structured patterns
having the symmetry of the underlying lattice, with the behavior
of agents belonging to a particular structure being highly correlated.
Note that these patterns are dynamic, being essentially concentric
waves that emerge at the center and travel to the  boundary of the region,
which continually expands until it meets another such pattern.
Where two patterns meet their progress is arrested and their
common boundary resembles a dislocation line.
In the asymptotic limit, several such patterns fill up the entire system.
Ordered patterns have previously been observed in spatial prisoner's
dilemma model \cite{nowak92}. However, in the present case, the patterns
indicate the growth of clusters with strictly correlated choice behavior.
The central site in these clusters act as the ``opinion leader'' for the
entire group.
This can be seen as analogous to the formation of ``cultural groups'' with
shared
beliefs \cite{axelrod97}.
It is of interest to note that distributing $\lambda$ from a random
distribution among the agents disrupt the symmetry of the patterns,
but we still observe patterns of correlated choice behavior (Fig.~\ref{fig:2}).
It is the global feedback ($\lambda \neq 0$)
which determines the formation of large connected regions of agents having
similar choice behavior.

\begin{figure}[tbp]
\centering
\includegraphics[height=6cm]{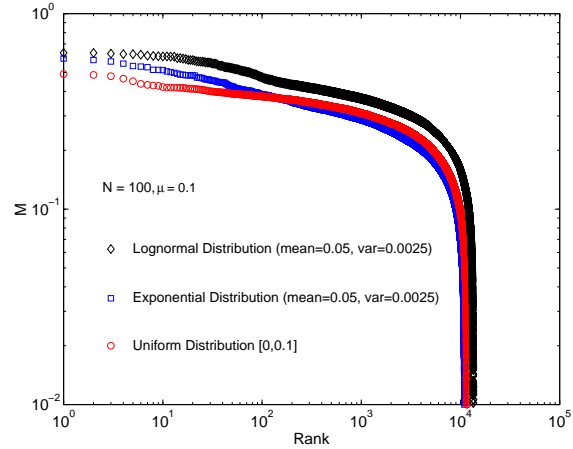}
\caption{Rank-ordered plot of $M$ for a one-dimensional lattice 
of $100$ agents. The adaptation rate $\mu = 0.1$, and the learning rate 
$\lambda$ of each agent
is chosen from three different random distributions: uniform (circle),
exponential (square) and log-normal (diamond).
}
\label{fig:5}       
\end{figure}

To get a better idea about the distribution of the magnitude of fractional
excess demand,
we have looked at the rank-ordered plot of $M$, i.e., the curve
obtained by putting the highest value of $M$ in position 1, the second highest
value of $M$ in position 2, and so on. As explained in Ref.~\cite{adamic02},
this plot is related to the cumulative distribution function of $M$.
The rank-ordering of $M$ shows that with $\lambda = 0$,
the distribution varies smoothly over a large range, while for $\lambda > 0$,
the largest values are close to each other, and then shows
a sudden decrease. In other words, the presence of global feedback results
in a high frequency of 
market events where the choice of a large number of agents become
coordinated, resulting in excess demand or supply.
Random distribution of $\lambda$ among the agents results in only small
changes to the curve (Fig.~\ref{fig:5}). 
However, the choice of certain distribution functions for $\lambda$
elevates the highest values of $M$ beyond the trend of the curve,
which reproduces an empirically observed feature in many popularity
distributions that has sometimes been referred to as
the ``king effect'' \cite{laherrere98,davies02}.

\section{Conclusion}
In summary, we have presented here a model for the emergence of collective
action defining market behavior through interactions between agents who
make decisions based on personal information that 
change over time through adaptation and learning. 
We find that introducing these effects produces market behavior marked
by two phases: (a) market equilibrium, where the buyers and sellers 
(and hence, demand and supply) are balanced, and (b) market polarization,
where either the buyers or the sellers dominate (resulting in excess
demand or excess supply).
There are multiple
mechanisms by which the transition to market polarization occurs, e.g., 
(i) keeping the adaptation and learning rate fixed but
switching from an initially regular neighborhood structure (lattice)
to a random structure (mean-field) one sees a transition from market
equilibrium to market polarization;
(ii) in the lattice, by increasing the learning rate
$\lambda$ (keeping $\mu$ fixed) one sees a transition from equilibrium to
polarization behavior; and (iii) in the case where agents have randomly chosen
neighbors, by increasing the adaptation rate $\mu$ beyond a critical value
(keeping $\lambda$ fixed) one sees a transition from polarized to equilibrium
market state.

The principal interesting observation seems to be that while, on the one hand,
individual agents regularly switch between alternate choices as a result
of adapting their beliefs in response to new information, on the other hand,
their collective action (and hence, the market) may remain polarized
in any one state for a prolonged period.
Apart from financial markets, such
phenomena has been observed, for example, in
voter behavior, where preferences have been observed to change
at the individual level which is not reflected in the collective level,
so that the same party remains in power for extended periods.
Similar behavior possibly underlies the emergence of cooperative behavior
in societies. As in our model, each agent can switch regularly between
cooperation and defection; however, society as a whole can get trapped in
a non-cooperative mode (or a cooperative mode) if there is a strong
global feedback.

Even with randomly distributed $\lambda$
we see qualitatively similar
results, which underlines their robustness.
In contrast to many current models, we have not assumed a priori
existence of contrarian and trend-follower strategies among the
agents \cite{lux95}.
Rather, such behavior emerges naturally from the micro-dynamics of
agents' choice behavior. Further, we have not considered external
information shocks, so that all observed fluctuations in market activity
is endogenous. This is supported by recent empirical studies which
have failed to observe any significant correlation between market
movements and exogenous economic variables like investment
climate \cite{kaizoji00}.

We have recently studied a variant of the model in which the 
degree of interactions between neighboring agents $J_{ij}$ is not uniform 
and static, but evolves in time \cite{sinha06b}. This is implemented by 
assuming that agents seek out the most successful
agents in its neighborhood, and choose to be influenced by them preferentially.
Here, {\em success} is measured by the fraction of time the agents decision 
(to buy or sell) accorded with the market behavior. The resulting model
exhibits extremely large fluctuations around the market equilibrium state
($M = 0$)
that quantitatively match the fluctuation distribution of stock price
(the ``inverse cubic law'') seen in real markets.

Another possible extension of the model involves introducing stochasticity in 
the dynamics. In real life, the information an agent obtains about
the choice behavior of other agents is not completely
reliable. This can
be incorporated in the model by making the updating 
rule Eq.~(\ref{seqn}) probabilistic.
The degree of randomness can be controlled by a ``temperature''
parameter, which
represents the degree of reliability an agent attaches to the information
available to it. Preliminary results indicate that higher temperature produces
unimodal distribution for the fractional excess demand.

Our results concerning the disparity between behavior at the level of the
individual agent, and that of a large group of such agents, has ramifications
beyond the immediate context of financial markets \cite{sinha06}. 
As for example, it is 
often said that ``democracies rarely go to war'' because getting a consensus 
about 
such a momentous event is difficult in a society where everyone's free opinion 
counts. This would indeed have been the case had it been true that the
decision of each agent is made independently of others, and is 
based upon all evidence available to it. However, such an argument
underestimates how much people are swayed by the collective opinion of those
around them, in addition to being aroused by demagoguery and yellow
journalism. Studying the harmless example of how market polarizations
occur even though individuals may regularly alternate between different
choices may help us in understanding 
how more dangerous mass madness-es can occur in a society.

\vspace{0.2cm}
\noindent
{\bf Acknowledgements}\\
We thank J.~Barkley Rosser, Bikas Chakrabarti, Deepak Dhar, Matteo Marsili, 
Mishael Milakovic, Ram Ramaswamy,
Purusattam Ray and Dietrich Stauffer for helpful discussions. SS would like
to thank the Santa Fe Institute where part of the work was done
and Sam Bowles, Jung-Kyoo Choi, Doyne Farmer and Lee Segel for comments.


\begin{thebibliography}{99}
\bibitem{farmer05} Farmer J D, Shubik M, Smith E (2005)
Is economics the next physical science~? Physics Today 58 (9):37--42

\bibitem{durlauf99} Durlauf S N (1999) How can statistical mechanics 
contribute to social science~? Proc. Natl. Acad. Sci. USA 96: 10582--10584

\bibitem{gopikrishnan98} Gopikrishnan P, Meyer M, Amaral L A N, 
Stanley H E (1998) Inverse cubic law
for the distribution of stock price variations, Eur. Phys. J. B 3:139--140

\bibitem{sinha06a} Sinha S, Pan R K (2006)
The power (law) of Indian markets: Analysing NSE and BSE trading statistics. 
In: Chatterjee A, Chakrabarti B K
(ed) Econophysics of stock and other markets. Springer, Milan

\bibitem{schiller00} Schiller R J (2000) Irrational exuberance.
Princeton University Press, Princeton

\bibitem{sinha04b} Sinha S, Raghavendra S (2004) Hollywood blockbusters and 
long-tailed distributions: An empirical study of the popularity of movies. 
Eur. Phys. J. B 42: 293--296

\bibitem{keynes34} Keynes J M (1934) The general theory of employment,
interest and money. Harcourt, New York

\bibitem{sinha04} Sinha S, Raghavendra S (2004)
Phase transition and pattern formation in a model of collective choice
dynamics. SFI Working Paper 04-09-028

\bibitem{iori02} Iori G (2002) A microsimulation of traders activity 
in the stock market: the role of heterogeneity, agents' interaction and 
trade frictions, J. Economic Behavior \& Organization 49:269--285

\bibitem{weisbuch03} Weisbuch G, Stauffer D (2003) 
Adjustment and social choice.
Physica A 323: 651--662

\bibitem{sinha05} Sinha S, Raghavendra S (2005)
Emergence of two-phase behavior in markets through interaction and learning
in agents with bounded rationality. In: Takayasu H (ed) Practical fruits
of econophysics. Springer, Tokyo :200--204

\bibitem{arthur89} Arthur B W (1989) Competing technologies, increasing 
returns, and lock-in by historical events. Economic J. 99: 116--131

\bibitem{nowak92} Nowak M A, May R M (1992) Evolutionary games and 
spatial chaos. Nature 359: 826--829

\bibitem{axelrod97} Axelrod R (1997) The dissemination of culture: 
A model with local convergence and global polarization. J. Conflict
Resolution 41: 203--226

\bibitem{adamic02} Adamic L A, Huberman B A (2002)
Zipf's law and the internet. Glottometrics 3:143--150

\bibitem{laherrere98} Laherrere J, Sornette D (1998) 
Stretched exponential distributions in nature and 
economy: ``fat tails'' with characteristic scales.
Eur. Phys. J. B 2: 525--539

\bibitem{davies02} Davies J A (2002) The individual success of musicians, 
like that of physicists, follows a stretched exponential distribution.
Eur. Phys. J. B 4: 445--447

\bibitem{lux95} Lux T (1995) Herd behaviour, bubbles and crashes.
Economic J. 105: 881--896

\bibitem{kaizoji00} Kaizoji T (2000) Speculative bubbles and crashes in 
stock markets: An interacting-agent model of speculative activity.
Physica A 287: 493--506

\bibitem{sinha06b} Sinha S (2006)
Apparent madness of crowds: Irrational collective behavior emerging 
from interactions among rational agents. In: Chatterjee A, Chakrabarti B K
(ed) Econophysics of stock and other markets. Springer, Milan

\bibitem{sinha06} Sinha S, Pan R K (2006)
How a ``hit'' is born: The emergence of popularity from the dynamics of
collective choice. In: Chatterjee A, Chakraborti A, Chakrabarti B K (eds)
Handbook of econophysics and sociophysics, Wiley-VCH

\end{thebibliography}
%



\printindex
\end{document}